# The local atomic quasicrystal structure of the icosahedral $Mg_{25}Y_{11}Zn_{64}$ alloy

S Brühne[1], E Uhrig[1], C Gross[1], W Assmus[1], A S Masadeh[2] and S J L Billinge[2]

[1] Physikalisches Institut, Johann Wolfgang Goethe-Universität, Robert-Mayer-Str. 2-4, D-60054 Frankfurt/Main, Germany
[2] Department of Physics and Astronomy and Center for Fundamental Materials Research, Michigan State University, East Lansing, MI 48824-1116, USA

**Abstract**
A local and medium range atomic structure model for the face centred icosahedral (*fci*) $Mg_{25}Y_{11}Zn_{64}$ alloy has been established in a sphere of $r = 27$ Å. The model was refined by least squares techniques using the atomic pair distribution (PDF) function obtained from synchrotron powder diffraction. Three hierarchies of the atomic arrangement can be found: (i) five types of local coordination polyhedra for the single atoms, four of which are of Frank-Kasper type. In turn, they (ii) form a three-shell (Bergman) cluster containing 104 atoms, which is condensed sharing its outer shell with its neighbouring clusters and (iii) a cluster connecting scheme corresponding to a three-dimensional tiling leaving space for few glue atoms. Inside adjacent clusters, $Y_8$-cubes are tilted with respect to each other and thus allow for overall icosahedral symmetry. It is shown that the title compound is essentially isomorphic to its holmium analogue. Therefore *fci*-Mg-Y-Zn can be seen as the representative structure type for the other rare earth analogues *fci*-Mg-Zn-*RE* (*RE* = Dy, Er, Ho, Tb) reported in the literature.

(Some figures in this article are in colour only in the electronic version)

## 1. Introduction

Since the discovery of icosahedral Al-Mn in 1984 [1] the determination of the atomic-scale structure of quasicrystals remains a difficult problem [2]. Another class of metastable iscosahedral alloys is Mg-Zn-based, containing Al or Ga. In 1993, Luo *et al* [3] discovered stable ternary Mg-Y-Zn quasicrystals with icosahedral diffraction symmetry. Y can also be substituted by Dy, Er, Gd, Ho and Tb [4]. Since 1998, single crystals have been available for Mg-Zn-*RE* (*RE* = Dy, Er, Ho, Tb) [5, 6]. The crystals exhibit a six-dimensional lattice parameter $a(6D) \approx 2 \times 5.2$ Å and an *F*-type centring called face-centred icosahedral (*fci*). *Fci*-Mg-Y-Zn shows virtually no diffuse scattering and therefore is considered to be of high structural perfection [7]. Mg-Zn-*RE* quasicrystals of comparable quality but with a *P*-type lattice, $a(6D) \approx 5.2$ Å are also known [8]. They are called simple icosahedral (*si*) and are found at higher zinc-magnesium ratios for *RE* = Er, Ho and Tm [9, 10].

Up to now, the following structure analyses of icosahedral Mg-Zn-*RE* quasicrystals are available: *fci*-Ho-Mg-Zn was refined in 6D and the derived 3D physical space information is given on the Ho-partial structure in the 2-fold plane [11]. Information on atomic clusters in *fci*-Mg-Tb-Zn was obtained from HRTEM and X-rays in [12] and a 6D Rietveld refinement of X-ray powder data revealed the average decoration of 3D Penrose tiles for *fci*-Ho-Mg-Zn [13]. Another 3D model for *fci*-Mg-Y-Zn was obtained by Fourier transform of single X-ray diffraction data [14]. In 2003 a quantitative analysis of the atomic pair distribution function (PDF) of *si*-Ho-Mg-Zn from in-house X-ray powder diffraction was performed and resulted in the element distribution and geometry of a 104-atom Bergman cluster [15]. The approach was based on rational approximant models for the local quasicrystal structure: While the 3D quasiperiodic structure can be generated from 6D *via* an irrational projection (using $\tau = (\sqrt{5}+1)/2$ in the projection matrix), a periodic *p/q* approximant is generated using *p/q* instead of $\tau$. As the PDF always reflects the *local* structure, the short range structure of the quasicrystal was refined *as if* it was a 1/1-approximant in a sphere confined to $r < 17$ Å [15]. A larger 2/1-approximant model contains eight such clusters and was similarily refined, in better agreement with the data, for *fci*-Ho-Mg-Zn with $r < 27$ Å [16]. Thus the PDF approach represents a complementary technique which just recently yielded detailed insight into the atomic structure of icosahedral Mg-Zn-*RE* phases.

In [16] we discussed the idea of "virtual" rational *p/q*-approximant models for the local structure of *fci*-Ho-Mg-Zn in detail. A cubic 2/1-approximant unit cell ($a \approx 23$ Å and symmetry restrictions *as if* in *Pa*-3 [18]) can serve as a coordinate system for the local model. In the present paper, that model has been adapted for the *fci*-Mg-Y-Zn phase. We use high $Q_{max}$ syncrotron powder diffraction data to generate a well resolved PDF for least-squares structure refinements. The questions to answer are: Can synchrotron data confirm our earlier results [16] from in-house X-rays? Is the Y-compound isostructural to the Ho-compound? If yes, this would open a perspective for future use of difference-PDFs since the PDF is a function on an absolute scale. Thus *local* atomic models containing only the *RE*-positions, *i.e.* only ~10% of all constituing atoms, could be used: Regarding the high unit cell contents (160, 680, 2888 or ~12200 atoms for 1/1, 2/1, 3/2 or 5/3 models, repectively [15]), this option would clearly simplify future structure calculations which are needed to understand the *quasiperiodic* structure.

## 2. Experimental details

*2.1 Synthesis and basic characterisation*

Single crystalline material of *fci*-Mg-Y-Zn has been obtained from the melt using the liquid-encapsulated top-seeded solution growth method (LETSSG) as described in [5]. Laue diffractograms of the grains exhibit symmetry *m*-3-5. The composition was determined to 64 at% Zn, 25 at% Mg and 11 at% Y by wavelength dispersive analysis of X-rays (WDX; Microspec WDX3PC) of polished samples within an accuracy of ± 1 at% *versus* standard specimens of the pure metals. Thus the formula for the title compound was chosen $Mg_{25}Y_{11}Zn_{64}$. Its density was determined to $\rho = 5.0(1)$ gcm$^{-3}$ measured by a He-pycnometer Micrometrics AccuPyc 1330. The X-ray powder diffractogram (Siemens Kristalloflex 810, CuK$\alpha$, $\lambda = 1.541$ Å) could be indexed with an *F*-centred lattice parameter $a(6D) \approx 2 \times 5.19(2)$ Å (reflection condition $h_1h_2h_3h_4h_5h_6$: $h_i$ all even or all odd) using Elser´s method [19; 9].

*2.2 Data collection*

The real-space pair distribution function (PDF), *G(r)*, gives the probability of finding pairs of atoms separated by distance *r*, and comprises peaks corresponding to all discrete interatomic distances. The experimental PDF is a direct Fourier transform of the total scattering structure function *S(Q)*, the corrected, normalized intensity, from powder scattering data given by

$$G(r) = \frac{2}{\pi} \int_0^\infty Q[S(Q)-1]\sin Qr\, dQ,$$

where $Q = \frac{4\pi}{\lambda}\sin\theta$ is the magnitude of the scattering vector. Unlike crystallographic techniques, the PDF incorporates both Bragg and diffuse scattering intensities resulting in local structural information [17, 20]. Its high real-space resolution is ensured by measurement of scattering intensities over an extended $Q$ range $Q_{max} \geq 35.0$ Å$^{-1}$ using short wavelength X-rays or neutrons.

The diffraction experiment was performed on a powdered sample at 6ID-D µCAT beamline at the Advance Photon Source (APS) at Argonne National Laboratory. Data acquisition at 300 K employed the recently developed rapid acquisition PDF (RA-PDF) technique [21] with an X-ray energy of 130.0 keV. Data were collected using an image plate camera (Mar345), with a usable diameter of 345mm, mounted orthogonal to the beam path with sample to detector distance of 220 mm. Lead shielding before the goniometer, with a small opening for the incident beam was used to reduce the background. All raw data were integrated using the software Fit2D [22] and converted to intensity versus 2θ (the angle between incident and the scattered X-rays). The integrated data were normalized with respect to the average monitor count, then transferred to the program PDFgetX2 [23] to carry out data reduction to obtain $S(Q)$ and the PDF $G(r)$ which are shown in Figure 1.

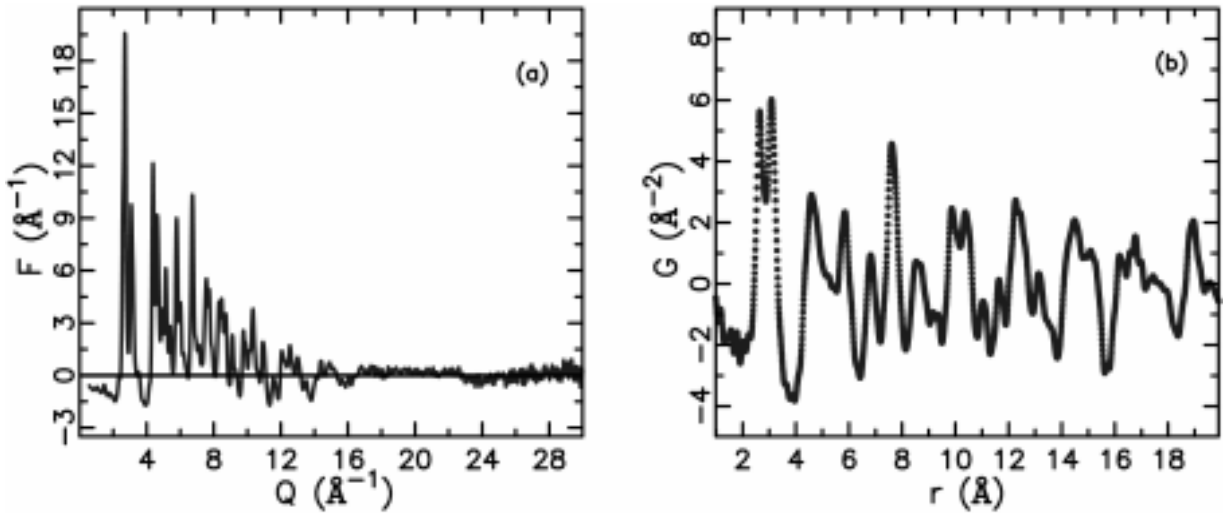

Figure 1. (a) The experimental reduced structure function $F(Q) = Q[S(Q)-1]$ of *fci*-Mg$_{25}$Y$_{11}$Zn$_{64}$ with $Q_{max}$ cut at 30.0 Å$^{-1}$ and (b) the corresponding PDF, $G(r)$.

*2.3 Structure refinements*

For the least squares structure refinement the program PDFFIT [24] was used. The starting model has the atomic coordinates of the "2/1"-model for the local structure of *fci*-Ho-Mg-Zn [16] where Ho was replaced by Y. The cubic lattice parameter was set to $a = 23.0$ Å and all temperature factors were set to $U_{eq} = 1.5\times10^{-2}$ Å$^{-2}$ for the beginning. The 832 data points in the *r*-range 2 to 27 Å were used for the refinements. This is a range of approximately 1.2 times the "virtual" lattice constant. Symmetry restrictions of *Pa*-3 [18] were retained for the atomic coordinates since this space group describes properly the "real" 2/1 approximants *e.g.* in the Al-Mg-Zn system. In a first step the scale factor, the dynamic correlation factor δ and lattice constant *a* were refined. Then the temperature factors $U_{eq}$ were allowed to relax and finally the positional parameters *xyz* were included in the refinements. Since $U_{eq}$ of Mg5 (nomenclature as in [16]) dropped to less than $0.01\times10^{-2}$ Å$^{-2}$, it was replaced by a zinc atom (Zn21). The refinements converge finally at $R$ = 13.7 %. The final plot of observed, calculated and difference data is given in Figure 2, the difference plot is almost featureless. $U_{eq}$ scatter statisticaly (see Table 2) – they "bury" the limitedness of the periodic model for an aperiodic structure. There is one exception (Y3) that is discussed in Sections 3.2 and 3.3. The resulting data were analyzed with respect to crystal-chemical validity [25]. Data concerning the refinement are listed in Table 1.

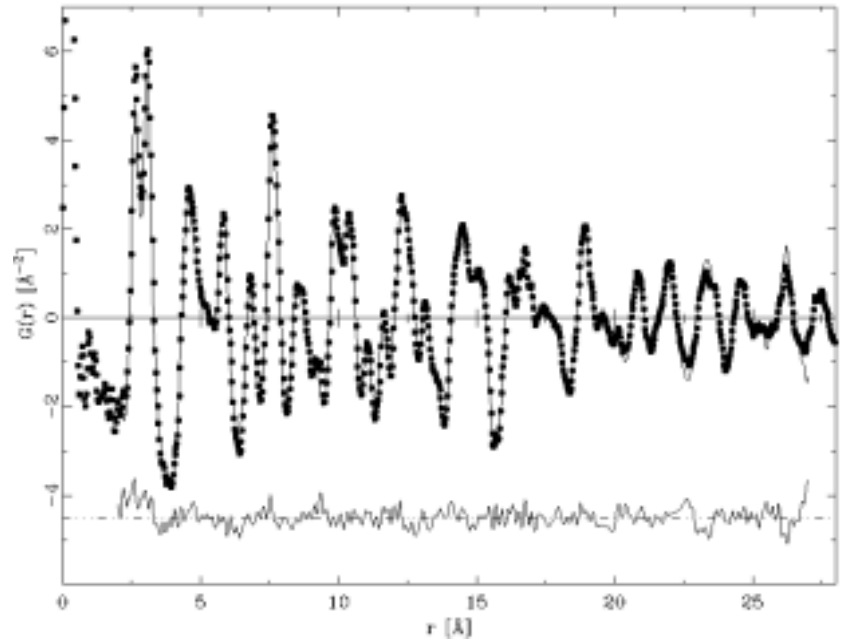

Figure 2. PDF $G(r)$ from synchrotron data of fci- $Mg_{25}Y_{11}Zn_{64}$ (dots), PDF calculated for the *local* atomic structure *as if* it was its cubic 2/1 approximant ($r_{max}$ = 27 Å, solid line) and their difference plot below ($R$ = 13.7 %).

Table 1. Data for the final least squares refinements of the *local* atomic structure of fci- $Mg_{25}Y_{11}Zn_{64}$ *as if* it was its cubic 2/1 approximant

| scale factor | 1.562(3) |
|---|---|
| dynamic correlation factor $\delta$/Å$^3$ | 0.5159(2) |
| low $r/\sigma$ ratio | 1.0 |
| virtual approximant space group [18] | $Pa$-3 (no. 205) |
| virtual approximant lattice parameter $a$(3D) /Å | 23.0291(5) |
| calculated hypercubic lattice parameter $a$(6D) /Å | 2×5.170 |
| refinement $r$ range in Å | 2 to 27 |
| number of data points used | 832 |
| $\lambda$/Å | 0.095 |
| termination at $Q_{max}$ /Å$^{-1}$ | 30.0 |
| calculated model composition | $Mg_{21.2}Y_{10.6}Zn_{68.2}$ |
| calculated model density $\rho$ / gcm$^{-3}$ | 5.436 |
| Q resolution $\sigma(Q)$ /Å-1 | 0.017 |
| number of refined parameters | 120 |
| $R$-value | 0.1371 |

## 3. Results and discussion

The structure refinements result in a data set for the local model for $fci$-$Mg_{25}Y_{11}Zn_{64}$ and is listed in Table 2. The data give rise to a structure description in terms of structural hierarchies. This will be discussed in Section 3.1 and 3.2. The Section 3.3 contains a comparison to $fci$-$Ho_9Mg_{26}Zn_{65}$.

Table 2. Structural data of $fci$-$Mg_{25}Y_{11}Zn_{64}$: The quasicrystal is described *locally as if* it was a 2/1-approximant, $a$ = 23.0291(5) Å with symmetry restrictions of space group $Pa$-3 (no. 205 [18]). The atom numbering scheme correponds to [16] for reasons of comparability. The average interatomic distance $\langle d \rangle$ is given for the first coordination shell with coordination number CN. Values in **bold** are discussed in the text.

| atom type | atom no. | Wyckoff position *as if* in $Pa$-3 | $x/a$ | $y/a$ | $z/a$ | $U_{eq}/10^{-2}\text{Å}^2$ | CN | $\langle d \rangle$/Å |
|---|---|---|---|---|---|---|---|---|
| $\alpha^0$ | [void] | 8$c$ | 0.3458 | $x$ | $x$ | [-] | 12 | 2.581 |
| $\alpha^1$ | Zn16 | 24$d$ | 0.2534 | 0.2912 | 0.3516 | 0.5 | 11 | 2.829 |
|  | Zn17 | 24$d$ | 0.2541 | 0.4098 | 0.3450 | 0.8 | 11 | 2.838 |
|  | Zn18 | 24$d$ | 0.2890 | 0.3464 | 0.4472 | 1.1 | 11 | 2.864 |
|  | Zn19 | 24$d$ | 0.2453 | 0.4408 | 0.4065 | 0.6 | 11 | 2.979 |
| $\alpha^2$ | Zn1 | 24$d$ | 0.0271 | 0.4712 | 0.1644 | 1.5 | 12 | 3.068 |
|  | Zn3 | 24$d$ | 0.0409 | 0.2302 | 0.1591 | 0.4 | 12 | 2.935 |
|  | Zn14 | 24$d$ | 0.1558 | 0.2389 | 0.3569 | 1.6 | 12 | 2.882 |
|  | Zn15 | 24$d$ | 0.1629 | 0.4642 | 0.3569 | 2.5 | 12 | 2.889 |
| $\alpha^3$ | Zn2 | 24$d$ | 0.0444 | 0.1314 | 0.1013 | 2.8 | 12 | 3.007 |
|  | Zn6 | 24$d$ | 0.0523 | 0.2896 | 0.3406 | 1.3 | 12 | 2.924 |
|  | Zn7 | 24$d$ | 0.0730 | 0.0911 | 0.4974 | 1.7 | 12 | 3.056 |
|  | Zn8 | 24$d$ | 0.0688 | 0.3914 | 0.3526 | 1.3 | 12 | 2.862 |
|  | Zn9 | 24$d$ | 0.0947 | 0.4674 | 0.4457 | 0.9 | 12 | 2.919 |
|  | Zn10 | 24$d$ | 0.0950 | 0.4699 | 0.2531 | 1.4 | 12 | 3.053 |
|  | Zn11 | 24$d$ | 0.1014 | 0.2303 | 0.2564 | 0.7 | 12 | 2.860 |
|  | Zn12 | 24$d$ | 0.1394 | 0.4044 | 0.1632 | 1.0 | 12 | 3.020 |
|  | Zn13 | 24$d$ | 0.1333 | 0.2887 | 0.1568 | 0.9 | 12 | 2.858 |
| $\beta$ | Mg3 | 24$d$ | 0.0536 | 0.3050 | 0.0666 | 1.0 | 16 | 3.352 |
|  | Mg7 | 24$d$ | 0.1622 | 0.3326 | 0.2775 | 0.6 | 16 | 3.184 |
|  | Mg8 | 24$d$ | 0.2261 | 0.2578 | 0.4666 | 0.3 | 16 | 3.206 |
|  | Mg9 | 24$d$ | 0.2132 | 0.4718 | 0.4397 | 3.7 | 16 | 3.242 |
|  | Y1 | 24$d$ | 0.0368 | 0.3505 | 0.2246 | 0.6 | 16 | 3.137 |
|  | Y2 | 24$d$ | 0.1496 | 0.3491 | 0.4221 | 2.0 | 16 | 3.155 |
|  | Y3 | 8$c$ | 0.2356 | $x$ | $x$ | **25.0** | 16 | 3.130 |
|  | Y4 | 8$c$ | 0.4609 | $x$ | $x$ | 0.2 | 16 | 3.324 |
| $\gamma$ | Zn5 | 24$d$ | 0.0560 | 0.1336 | 0.2373 | 1.8 | 14 | 3.010 |
|  | Zn21 | 24$d$ | 0.0992 | 0.2142 | 0.4553 | 4.4 | 14 | 3.124 |
| $\delta^X$ | Mg1 | 24$d$ | 0.0332 | 0.3544 | 0.4620 | 1.6 | 15 | 3.131 |
|  | Mg4 | 24$d$ | 0.0499 | 0.1584 | 0.3580 | 2.3 | 15 | 3.064 |
| $\delta^Y$ | Y5 | 8$c$ | 0.1581 | $x$ | $x$ | 1.3 | 16 | 3.071 |
| $\delta^Z$ | Zn20 | 8$c$ | 0,0220 | $x$ | $x$ | 3.3 | **12** | 2.957 |

*3.1 Structure description*

Three structural hierachies are observed on different length scales in *fci*-Mg-Y-Zn:
(i) Local atomic coordination polyhedra ($r < 4$ Å) all exhibit coordination numbers (CN) 11, 12, 14, 15 or 16 with the same topologies, respectively.
(ii) The atoms of (i) group to units of 104 atoms (Bergman cluster, $r \approx 15$ Å).
(iii) These clusters are arranged on the vertices of a Canonical Cell Tiling (CCT, [29]). There are only two tiling edge lengths: ~12 Å and ~14 Å.

Note that due to the self-similar character of quasicrystalline structures the icosahedral topology will be observed at different points: as local corrdination polyhedra (CN 12, to be found all over the structure), in shell 1 and 2b of the cluster and finally as overall diffraction symmetry. The three structural hierarchies (i) to (iii) will be addressed below:

*ad (i):* Except CN 11, the coordination shells are all triangulated and they topologically represent regular, or sometimes distorted, Frank-Kaspar polyhedra [26, 27]. According to their metallic radii [28], Zn atoms reside in CN 11, 12, and 14; Mg atoms in CN 15 or 16; all Y atoms require CN 16. Most frequent is the icosahedron (CN 12) for 47 % of all atoms. CN 11, however, can be described as CN (12-1) since it is topologically an icosahedron that lacks one vertex. For typical examples of the five types of polyhedra, see Figure 3.

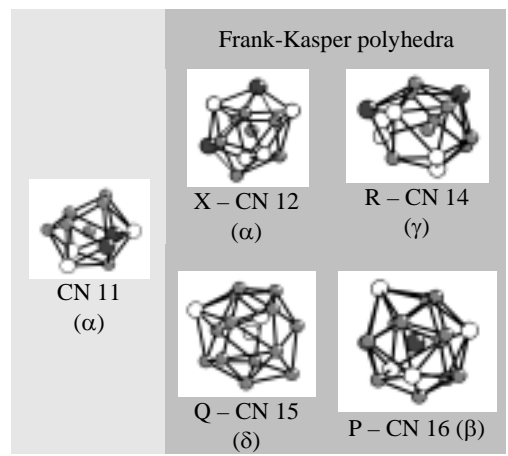

Figure 3: Five types of local coordination polyhedra in *fci*-Mg-Y-Zn. White balls: Mg atoms; grey: Zn; dark grey: Y. Capital letters denote Frank-Kasper polyherdra [26]; CN: coordination number (greek letters refer to structural function of the central atom on the next hierarchy level).

*ad (ii):* To define the structural function of the single atoms at the second hierarchical level, they are labelled after [30] using Greek letters: $\alpha^1$, $\alpha^2$, $\alpha^3$, $\beta$, $\gamma$ and $\delta$. Figure 4 explains the architecture of the cluster. It is built of three concentric shells:
(1) 12 Zn atoms ($\alpha^1$) are placed around the void cluster centre ($\alpha^0$) forming an empty icosahedron of $r \approx 3$ Å.
(2a) 12 Mg and 8 Y atoms ($\beta$) form a pentagon-dodecahedron. Its ideal (topological) symmetry *m*-3-5 is lowered to *m*-3 since the 8 Y atoms form an inscribed cube of edge length 5.4 Å.
(2b) The second shell is completed by 12 Zn atoms ($\alpha^2$) which lay on the vertices of a $\tau$-inflated analogue of the first shell (icosahedron).
All atoms of shell (2) represent a rhombic dodecahedron of $r \approx 4$ to 5.5 Å.
(3) The third shell consists of 48 $\alpha^3$ atoms and 12 $\gamma$ atoms, in total 60 Zn atoms arranged like a truncated icosahedron or soccer ball of $r \approx 15$ Å.

While the innermost shell (1) is an empty regular icosahedron, the outer shells are more distorted due to interaction with neighbouring clusters, see (iii).

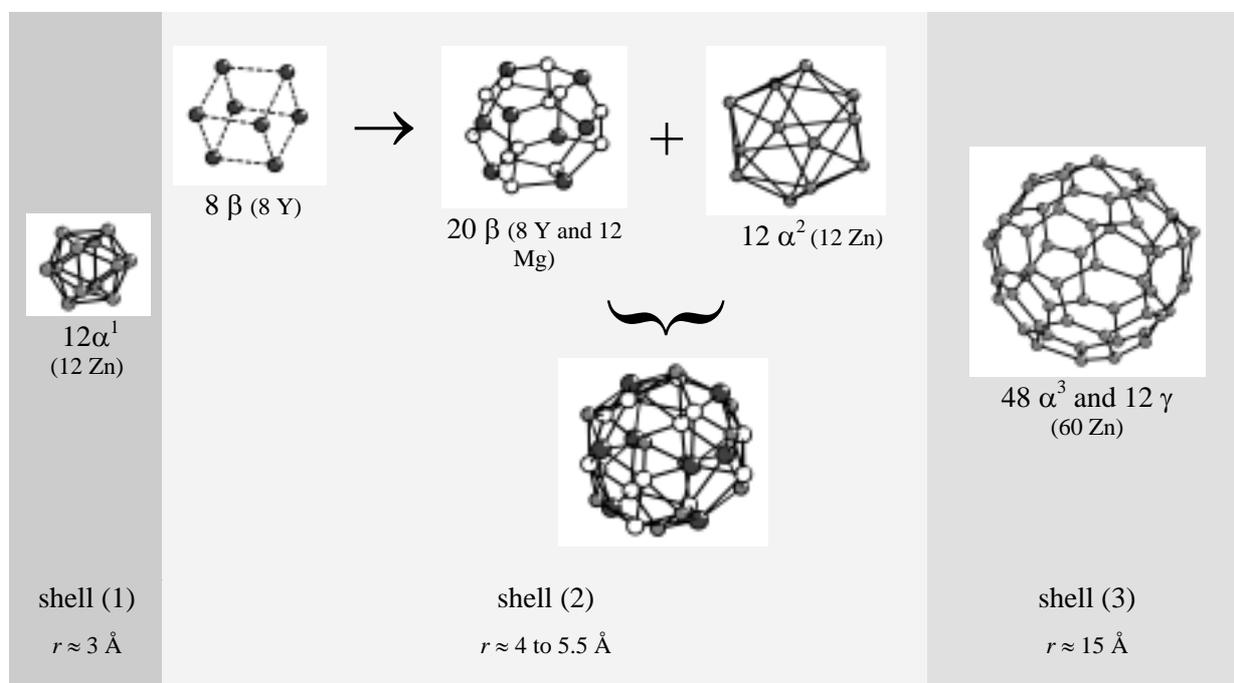

Figure 4: Three concentric shells together build the basic, onion-like structural unit in *fci*-$Mg_{25}Y_{11}Zn_{64}$ (Bergman cluster, in total 104 atoms). White balls: Mg atoms; grey: Zn; dark grey: Y. 1st and 3rd shells consist of Zn atoms only, shell (2) shows a distinct distribution of Y, Mg and Zn; 8 Y atoms are arranged on the vertices of a cube.

*ad (iii):* The $\alpha^3$ atoms are shared with the neighbouring clusters and define a so-called "c-bond" of the CCT. $\gamma$ atoms are shared as well and define a "b-bond" of the CCT. The CCT consists of four types of 3D cells (namely A, B, C and D) made of three faces (termed X, Y and Z) which all consist of b- and c-bonds; see Figure 5.

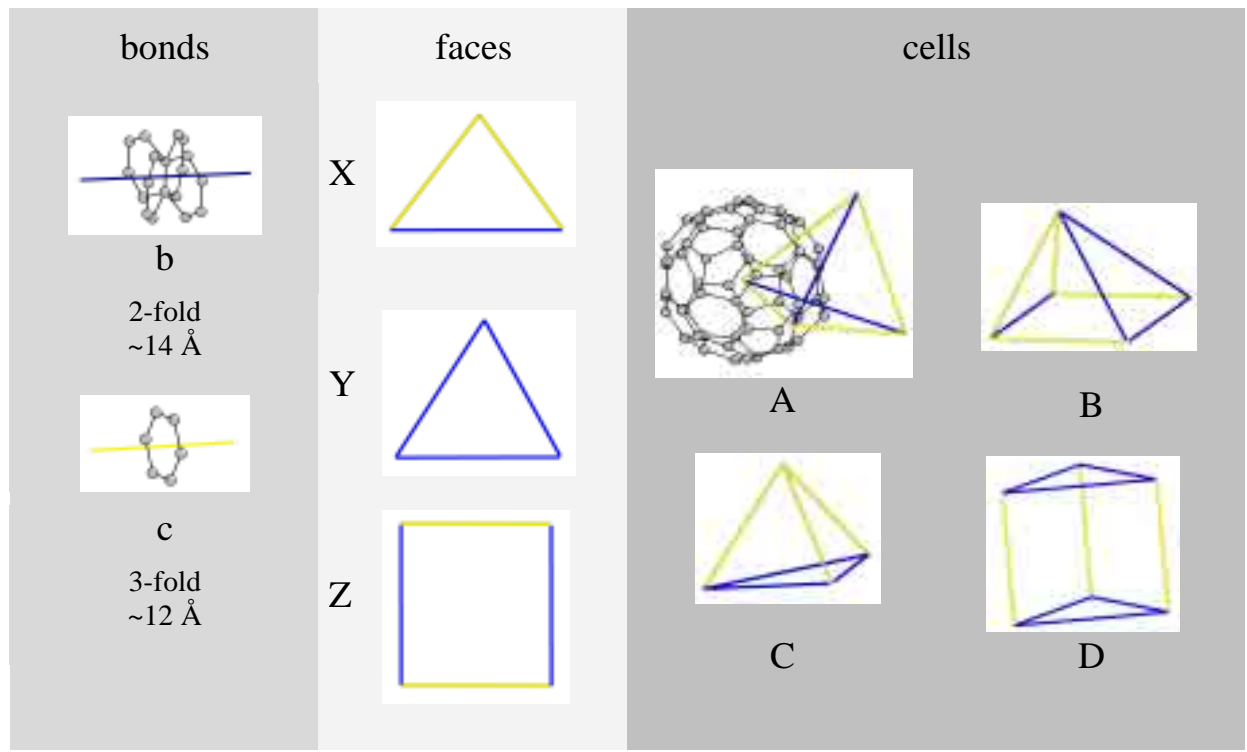

Figure 5: Bonds, faces and cells of the Canonical Cell Tiling (CCT [29]). b-bonds (c-bonds) connect icosahedral objects along their 2-fold (3-fold) symmetry directions. It is $|c| = (\sqrt{3}/2) \times |b|$. Some atoms of the third shell are drawn in to enlight the context.

In our local 2/1-model each node of the CCT represents a cluster centre. At each node six b-bonds and seven c-bonds meet in such a way that nine A, three B and three C-cells fill the whole space around the node as shown in Figure 6. This node environment represents the local matching rule for the local model.

In between the clusters, space is filled by so called glue atoms ($\delta$). Their location is always in the plane of a CCT face, so $\delta^X$, $\delta^Y$ and $\delta^Z$ atoms are distinguished (see Table 2). Both the common $\alpha^3$ and $\gamma$ atoms in the third shells and the stuffing with $\delta$ atoms implies a close packed stucture. Therefore, the term "cluster" as it is used here has not to be mistaken for isolated clusters, see also [2].

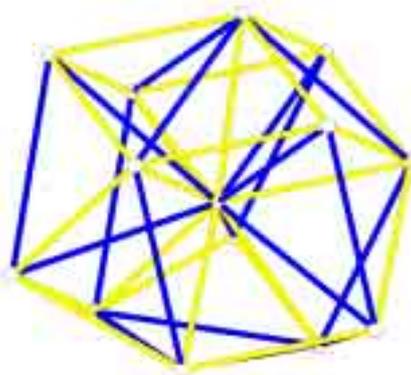

Figure 6: CCT node environment (local matching rule) for the 2/1 approximant local structure model for *fci*-$Mg_{25}Y_{11}Zn_{64}$.

*3.2 Details of the local model for* fci-*Mg-Y-Zn*

The crystal-chemical validity of the proposed model can be considered by plotting the average distances of central atoms to their coordinating ligands, $\langle d \rangle$. A plot of $\langle d \rangle$ *vs.* coordination number CN is shown in Figure 7. The values make chemical sense for the metallic radii present in the structure [28]. There is also a clear trend of increasing $\langle d \rangle$ at higher CN as expected. The excellent agreement of the model with the measured PDF (Fig. 2) also indicates that the model yields the correct real-space local structure. The outlier at ($\langle d \rangle$ = 2.581 Å, CN = 12) corresponds to the void at the cluster centre. There is not enough space to accommodate a hypothetical Zn atom which is consistent with this site being a vacancy.

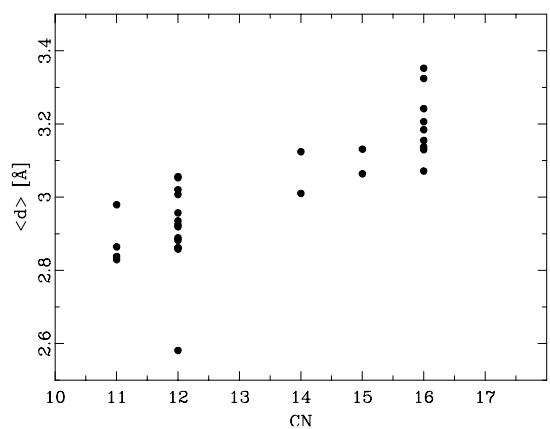

Figure 7. Plot of average interatomic distances $\langle d \rangle$ *vs.* coordination number CN in the local model for *fci*-Mg-Y-Zn

There is one unphysically short interatomic distance in the model: $d$(Zn20-Zn20) = 1.76 Å. Zn20 is a $\delta^Z$ glue atom that lies in the rectangular Z face of the CCT. Counting the short distance, the CN would be 13. Instead we can regard it as a split position. A regular CN 12 polyhedron results and the anoomalously short Zn-Zn interaction is removed. For this reason we choose the latter scenario and the coordination presented in Table 2 reflects this.

The calculated model composition is $Mg_{21.2}Y_{10.6}Zn_{68.2}$. This is somewhat richer in zinc than the measured composition $Mg_{25}Y_{11}Zn_{64}$. The calculated density $\rho_{calc} = 5.436$ gcm$^{-3}$ compares to the measured value $\rho_{meas} = 5.0(1)$ gcm$^{-3}$. These observations can be accounted for if we assume that some Mg sits on the γ positions, there is a lower occupancy of $\delta^Z$ (Zn20) and the measured density is an underestimation of the fully dense material due to the possible presence of pores in the sample. For example, the assumption all γ positions being totally occupied by Mg would lead to $\rho_{calc} = 5.129$ gcm$^{-3}$.

The Y partial structure consists of $Y_8$-cubes (Y1 to Y4) of edge length 5.4 Å which are tilted with respect to each other in four of the five different possible orientations to inscribe a cube in a pentagon dodecahedron. In that way global icosahedral symmetry can be achieved in the quasicrystal. Figure 8 explains the interconnection of the cubes *via* the $\delta^Y$ glue atom, Y5, attached to a canonical C cell. Here equilateral $Y_3$-triangles with an edge length of again 5.4 Å occur in a twisted manner around Y5. In [16] sterical reasons are brought forward as an argument for the absence of direct *RE-RE* contacts < 5 Å. This is also consistent with findings from an EXAFS investigation [31] and may explain the high temperature factor $U_{eq}(Y3) = 0.25$ Å$^2$ (see Table 2) since Y3 is connected to Y5 at $d = 3.1$ Å in our model.

The details discussed concerning composition, density and especially the properties of Zn20 and Y5 all touch the question for the 'true' cluster connection scheme in the quasicrystal. Beyond the local model developed here, there is evidence for an interpenetration of some of the clusters in higher approximant structures (*e.g.* 3/2-2/1-2/1-Ga-Mg-Zn, [30]) or other models for *fci*-Mg-Zn-*RE* [32, 14]. Unfortunately, an icosahedral quasiperiodic CCT has not been found by mathematicians yet [33]. As is the case for other tiling approaches, there is an intrinsic interdependence between the atomic decoration and the tiles themselves [32]. PDF quasicrystal analysis using an *r*-range and a model both confined to ~25 Å cannot give a satisfactory answer to that question.

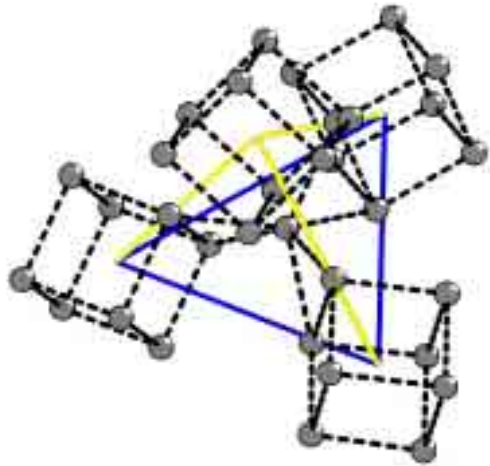

Figure 8. Y partial structure in *fci*-$Mg_{25}Y_{11}Zn_{64}$ around the four nodes of a canonical C cell. The $Y_8$ cubes are connected *via* a central $\delta^Y$ glue Y atom. Dotted interatomic spacings Y-Y: 5.4 Å.

*3.3 Relation to* fci-*Ho-Mg-Zn*

The diffraction patterns of *fci*-Mg-Y-Zn and *fci*-Ho-Mg-Zn are very similar. Differences in the intensity distribution hitherto were assumed to be due to the different scattering powers of $^{39}$Y and $^{67}$Ho. Now we can confirm that the short-range atomic structure of *fci*-Mg$_{25}$Y$_{11}$Zn$_{64}$ is basically identical to that of *fci*-Ho$_9$Mg$_{26}$Zn$_{65}$ [16], both referred to as *fci*-Mg-Zn-*RE* (*RE* = Y; Ho). Compared to our earlier study on the Ho compound we find the following comparisons:

Displacements of the fractional atomic coordinates with respect to *a*(3D) of the cubic 2/1 model cells are 0.01 at average and 0.04 at maximum; this means 0.3 and 0.9 Å, respectively, on an absolute scale. The cluster centers ($\alpha^0$) in both phases are not occupied and the $\alpha^1$ atoms form a regular icosahedron. The element distribution Y/Mg on the β positions corresponds to the Ho/Mg distribution (*RE*$_8$ cubes). The large temperature factor $U_{eq}$ of Y3 is also mirrored in the same trend for $U_{eq}$(Ho3) in [16]. Regarding the *RE* partial structure in *fci*-Mg-Zn-*RE*, here is a limit of the 2/1 approximant local structure model and points to a more complicated cluster connecting scheme in the "true" quasicrystal structure. $\alpha^2$ and $\alpha^3$ atoms practically coincide in both *fci* phases. A difference is observed for the γ atoms: Whereas in *fci*-Ho-Mg-Zn the Zn5 position tends to be occupied by Mg, in the *fci*-Mg-Y-Zn homologue Mg5 had to be replaced by Zn21. CN 14 allows for both elements at the γ position, whereas it is completely occupied by Zn in the zinc rich compound *si*-Ho$_{11}$Mg$_{15}$Zn$_{74}$ [15]. For the glue atoms, coincidence is found for $\delta^X$ (Mg1 and Mg4) and $\delta^Y$ (Y5 substitutes for Ho5) positions. A difference is visible at the coordination of the $\delta^Z$ position (Zn20): In *fci*-Ho-Mg-Zn it results in CN 13 – on the other hand in *fci*-Mg-Y-Zn the short distance *d*(Zn20-Zn20) rises the question whether there is a split position (resulting in CN 12) or whether there exists a CCT rectangular Z face in the real quasicrystal at all?

## 4. Conclusion

This structural investigation basing on synchrotron powder diffraction data and PDF analysis ($Q_{max}$ = 30 Å$^{-1}$) of *fci*-Mg-Y-Zn compares to the earlier in-house result for *fci*-Ho-Mg-Zn ($Q_{max}$ = 13.5 Å$^{-1}$, [16]) quite nicely: PDF refinements for *fci*-Mg$_{25}$Y$_{11}$Zn$_{64}$ (*R* = 13.7 %) confirm the local cluster architecture. Both *fci* phases show basically the same topological features and element distribution, Y substitutes for Ho in the respective partial *RE* (*RE* = Y, Ho) structures. A generic feature is the *RE*$_8$ cube (edge length 5.4 Å) inscribed in the second shell of the Bergman cluster. Minor differences in between *fci*-Mg-Y-Zn and *fci*-Ho-Mg-Zn point to the limit of the local model: the "true" cluster connection scheme in the quasicrystal is more complicated, see also [14, 32].

To resolve this problem, large models which contain the *RE* partial structure only (~10 % of all atoms in the alloy), should be accessible using difference-PDFs $\Delta G(r)_{RE} = G(r)_{fci\text{-Ho-Mg-Zn}} - G(r)_{fci\text{-Mg-Y-Zn}}$ since the substitution of Y by Ho is seen to be isomorphic. This will be subject of a future publication.

Nevertheless, the refined model that is presented here, will be found locally in the icosahedral quasicrystal structure of *fci*-Mg-Zn-*RE* alloys. *Fci*-Zn-Y-Mg can be seen as the representative structure type for the other rare earth analogues *fci*-Mg-Zn-*RE* (*RE* = Dy, Er, Gd, Ho, Tb) reported in the literature.

**Acknowledgements**
SJB would like to acknowledge help from Didier Wermeille and Doug Robinson for help in collecting data. Work in the Billinge-group was supported by Department of Energy (DOE) grant DE-FG02-97ER45651. Data were collected at the 6IDD beamline in the Midwest Universities Collaborative Access Team (MUCAT) sector at the Advanced Photon Source (APS). Use of the APS is supported by the U.S. DOE, Office of Science, Office of Basic Energy Sciences, under Contract No. W-31-109-Eng-38. The MUCAT sector at the APS is supported by the U.S. DOE, Office of Science, Office of Basic Energy Sciences, through the Ames Laboratory under Contract No. W-7405-Eng-82.